\documentclass[preprint,11pt]{elsarticle}

\usepackage{amssymb}
\usepackage{tabularx,ragged2e,booktabs,caption}
\usepackage{amsmath,amssymb,amsfonts}
\usepackage{algorithmic}
\usepackage{graphicx}
\usepackage{textcomp}
\usepackage{xcolor,soul}
\usepackage[english]{babel}
\usepackage{amsthm}
\usepackage[english]{babel}
\usepackage{verbatim}
\usepackage{comment}
\newtheorem{theorem}{Theorem}[section]

\newtheorem{lemma}{Lemma}[section]
\newtheorem{assumption}{Assumption}

\newtheorem{definition}{Definition}[section]

\usepackage[ruled]{algorithm2e}
\usepackage{ulem}
\usepackage{pstricks}
\usepackage{url}
\usepackage{bm}

\usepackage{amssymb}
\usepackage{amsmath}

\newcommand{\bmat}[1]{\begin{bmatrix} #1 \end{bmatrix}}

\journal{Applied Energy}

\begin{document}

\begin{frontmatter}



\title{Peer-to-Peer Energy Trading of Solar and Energy Storage: A
Networked Multiagent Reinforcement Learning Approach}


\author[label1]{Chen Feng} 
\author[label1]{Andrew L. Liu\corref{cor1}}
\affiliation[label1]{organization={School of Industrial Engineering, Purdue University},
            addressline={Grissom Hall}, 
            city={West Lafayette},
            postcode={47907}, 
            state={IN},
            country={USA}}

\cortext[cor1]{Corresponding author. E-mail address: andrewliu@purdue.edu.}
\begin{abstract}
Utilizing distributed renewable energy resources, particularly solar and energy storage, in local distribution networks via peer-to-peer (P2P) energy trading has long been touted as a solution to improve energy systems' resilience and
sustainability. Consumers and prosumers (that is, those with solar PV and/or energy storage), however, do not have the expertise to engage in
repeated P2P trading, and the zero-marginal costs of renewables
present challenges in determining fair market prices. To address
these issues, we propose multi-agent reinforcement learning
(MARL) frameworks to help automate consumers’
bidding and management of their solar PV and energy storage
resources, under a specific P2P clearing mechanism that utilizes
the so-called supply-demand ratio. In addition, we show how
the MARL frameworks can integrate physical network
constraints to realize voltage control, hence ensuring physical
feasibility of the P2P energy trading and paving way for real-world implementations.
\end{abstract}




\end{frontmatter}



\section{Introduction}\label{sec:Intro}
As our society strives to transition towards sustainable energy sources, distributed renewable resources and energy storage are increasingly seen as key components in creating resilient and sustainable energy systems. Peer-to-peer (P2P) energy trading in local energy markets offers a promising approach to decentralize and optimize the allocation of distributed energy resources (DERs). This model not only empowers consumers and those who can store or generate energy, known as prosumers, to trade energy directly but also promotes renewable energy usage and reduces energy losses.

However, implementing P2P trading presents significant challenges. First, consumers and prosumers lack the technical expertise required to engage in repeated P2P trading and manage their energy resources efficiently. Second, renewable energy, such as solar, often has zero marginal costs, which introduces difficulties in determining fair market prices for energy trades: a uniform-pricing auction resembling a wholesale energy market would not work since the market clearing prices would be zero most of the time. Third, while P2P trading only entails financial transactions, the delivery of energy occurs over the physical distribution networks. How to maintain network feasibility in a local P2P trading market is an important yet open question. 

In response to these challenges, this paper introduces a decentralized framework utilizing multiagent reinforcement learning (MARL), which is designed to automate the bidding processes of agents, enhancing the management efficiency of their solar PV and energy storage resources. Additionally, the framework ensures the feasibility of a distribution network. Specifically, inspired by the algorithm in \cite{zhang2018networked}, we propose a consensus-based actor-critic algorithm for a decentralized MARL, in which each agent in a repeated P2P auction is modeled as a Markov decision problem (MDP). By allowing agents to exchange information and reach agreements, the proposed framework facilitates efficient decision-making and resource allocation, while mitigating the computational and privacy challenges associated with certain centralized learning approaches. 
In addition, the shared network constraints, such as voltage regulation, can be learned through the decentralized approach with a constraint violation incurring a (fictitious) penalty to each agent's reward function. We consider this development a significant step towards practical real-world implementations of P2P energy trading.

In addition to the modeling and algorithm development, we establish the theoretical conditions for the consensus-MARL algorithm to converge to an asymptotically stable equilibrium. We consider this a significant advantage over a purely decentralized approach without agents' communication, such as the one in our previous work \cite{feng2022decentralized}, in which each agent uses the proximal policy optimization (PPO) approach \cite{PPO} to solve their own MDP and ignores multiagent interaction.  

Through numerical simulations, we compare three different frameworks to implement P2P trading while considering network constraints: purely decentralized trading using PPO, 
consensus-MARL, and a centralized learning and decentralized execution approach using the multiagent deep deterministic policy gradient (MADDPG) algorithm \cite{MADDPG}. Our findings indicate that consensus-MARL achieves higher average rewards compared to both PPO and MADDPG. 

The rest of the paper is organized as follows. Section \ref{sec:LitReview} reviews the literature on P2P energy trading and emphasizes the contribution of our work. Section \ref{sec:market} introduces the market clearing mechanism based on the supply-demand ratio, first proposed in \cite{liu2017energy}, which can avoid zero-clearing prices when the supply resources have zero marginal costs. Section \ref{sec:Consensus} provides the detailed consensus-MARL formulation and algorithm, as well as establishing convergence results. The specific setups and data of our simulation are described in Section \ref{sec:Sim}, along with numerical results. Finally, Section \ref{sec:Conclusion} summarizes our work and points out several future research directions. 

\section{Literature Review}
\label{sec:LitReview}

The literature on decentralized control of energy transactions in a distribution network is extensive and diverse. Broadly speaking, there are two predominant approaches: the distributed optimization approach and P2P bilateral trading. A key method in the former approach is the alternating direction multiplier method (ADMM), which solves centralized dispatch problems by iteratively updating price signals for consumer/prosumer optimization (such as  \cite{Multiclass,FADMM,liu2022fully,P2PADMM, IC_P2P}). Despite ADMM's theoretical convergence under convexity, practical challenges arise:
(i) As an algorithmic approach rather than a market
design, ADMM is subject to the same difficulties faced by
a uniform-pricing-based market of all zero-marginal-cost
resources. Specifically, it is prone to ‘bang-bang’ pricing: market clearing prices plummet to zero in oversupply
or spike to a price ceiling in under-supply. Addressing this
with additional market mechanisms, such as adding ancillary service markets, akin to wholesale markets,
could overly complicate distribution-level markets for energy
trading.
(ii) Most distributed algorithms, including ADMM, are static, lacking real-time adaptability. While online ADMM variants exist \cite{OnlineADMM,DualAscent}, ensuring network constraints within each iteration remains an open question.
(iii) Assuming consumers or prosumers can solve complex optimization problems is often unrealistic due to potential expertise or resource limitations.

These limitations form the basis of our interest in exploring P2P bilateral trading. Consequently, our literature review will focus on this alternative approach. There have been multiple review papers on the large and ever-growing literature on P2P energy trading \cite{PoorP2PReview,PoorP2PReview2,PoorGameTheory,P2PReview1,P2PReview23}.  Within the literature, we focus on the works that explicitly consider distribution network constraints. \cite{ZIP} proposes a continuous double auction framework within which physical network constraints are maintained through a sensitivity-based approach. The agents in the auction are the so-called zero-intelligence-plus (ZIP) traders, who employ simple adaptive mechanisms without any learning of the repeated multi-agent interactions.
More sophisticated than ZIP,  a fully decentralized MARL is adopted in \cite{biagioni2021powergridworld} to realize decentralized voltage control and is extended in \cite{feng2022decentralized} to form a P2P energy trading market with physical constraints. These two papers implement a completely decentralized approach in the sense that each agent ignores the multi-agent interaction as well and just learns their single-agent policy to maximize their own payoff. 

Another MARL approach that has been explored is the centralized training and decentralized execution (CTDE) framework, of which the multi-agent deep deterministic
policy gradient (MADDPG) method \cite{MADDPG} is a prominent example.
It is applied in \cite{wang2020data} to solve the autonomous voltage control problem 
In \cite{qiu2021multi}, a modified MADDPG algorithm is proposed for a double-sided auction market to facilitate P2P energy trading, although physical constraints are not considered in their model. While the CTDE framework can lead to more efficient learning and robust outcomes, its dependence on a centralized entity for collecting and coordinating all agents' data introduces difficulties for practical implementations and potential vulnerabilities, such as increased risk of single-point failures and higher susceptibility to cyber-attacks.

To overcome the drawbacks of fully decentralized and CTDE-based MARL, algorithms with communication among agents have also been developed. A notable contribution in this area is the consensus-based MARL algorithm designed for discrete-space problems, as presented in \cite{MARLBasar}, and its extension to continuous-space problems in \cite{zhang2018networked}. This model allows agents to learn and implement policies in a decentralized manner with limited communication, underpinned by a theoretical guarantee of convergence under certain conditions. Separately, \cite{qu2022scalable} introduced a Scalable Actor Critic (SAC) framework, particularly relevant for large-scale networks where an individual agent's state transition is influenced only by its neighbors. SAC is distinguished by its reduced communication needs compared to the CTDE framework and offers a theoretical guarantee for its policy convergence as well. However, the SAC framework is only established for discrete state space problems, and its extension and effectiveness in continuous-state space applications remain unexplored.

In this work,  we present several key contributions: First, we develop a comprehensive framework for a P2P energy trading market while integrating a consensus-based MARL algorithm. This framework facilitates automated control and bidding of DERs while allowing for decentralized learning of voltage constraints. Second, we present the sufficient conditions required for the convergence of the MARL algorithm and show 
the details of implementing this algorithm using deep neural networks. Third, we conduct a comparative analysis of market outcomes derived from three distinct MARL strategies: a naive decentralization approach, the MADDPG approach, and the proposed consensus-based MARL. Our findings reveal that the consensus-based approach notably outperforms the others in achieving better market outcomes. Our work addresses real-world applications of P2P energy trading under network constraints, which is crucial for integrating distributed renewable energy sources and enhancing grid stability. The RL-based algorithm we propose, designed for control automation, can be directly implemented on grid-edge devices, making it highly applicable for distribution networks with increasing numbers of prosumers.

\section{Market Clearing Mechanism}
\label{sec:market}
In this section, we first describe a clearing mechanism, referred to as the supply-demand-ratio (SDR)\cite{liu2017energy}, that can address the two potential issues faced by a renewable-dominated P2P market: (i) marginal-cost-based pricing yielding zero market prices, and (ii) the bang-bang outcomes in a double-auction-based mechanism, as the reserve prices for buyers (the utility rates) and the sellers (usually the feed-in tariffs) are publicly known. The second issue is unique to the P2P energy market, in our opinion, and the bang-bang phenomenon is documented in \cite{zhao2021auction} and will be explained further in this section. 

The SDR-clearing mechanism in P2P markets has been detailed in our earlier work \cite{feng2022decentralized}. For completeness, we briefly revisit it here. Let $\mathcal{I} = \{1,2,...,I\}$ represent all consumers and prosumers, all referred to as agents. We assume that trading among the agents takes place in fixed rounds (such as hourly or every 15 minutes) and is cleared ahead of time (such as day-ahead or hour-ahead). In each trading round $t$, each agent $i \in \mathcal{I}$ submits bids ($b_{i,t}$) to buy (if $b_{i,t} < 0$) or sell (if $b_{i,t} > 0$) energy. Agents can switch roles between buyers and sellers across different rounds, but not within the same round. For a round $t$, the sets of buyers and sellers are denoted as $\mathcal{B}_t = \{i: b_{i,t} < 0\}$ and $\mathcal{S}_t = \{i: b_{i,t} \geq 0\}$, respectively.

The SDR mechanism, originally proposed in \cite{liu2017energy}, is an approach to clear a market with all zero-marginal-cost supply resources, which is straightforward to implement. An SDR is defined as the ratio between total supply and demand bids at $t$; that is,
\vspace*{-8pt}
\begin{align}
	SDR_t := \displaystyle \sum_{i \in \mathcal{S}_t} b_{i,t}\bigg/\displaystyle -\sum_{i \in \mathcal{B}_t} b_{i,t}. 
	\label{eq:SDR}
\end{align}\\[-10pt]
To ensure that $SDR_t$ is well-defined, we assume $\mathcal{B}_t \neq \emptyset$ for any $t$, a reasonable assumption given agents' continuous need for energy. In rare cases where every agent is a prosumer with excess energy at time $t$, the P2P market can be temporarily suspended, and surplus energy sold to the grid at a predefined price, which will not impact our model or algorithm framework. Conversely, if $\mathcal{S}_t$ is $\emptyset$, indicating no excess energy for sale at $t$, then $SDR_t = 0$.

Based on its definition, when $SDR_t > 1$, it means that there is an over-supply. In this case, we assume that the excess energy in the P2P market is sold to a utility company, a distribution system operator (DSO), or an aggregator at a pre-defined rate, generally denoted as $FIT$ (feed-in tariff).\footnote{The $FIT$ rate is a rate set by utilities/policymakers. It can be set as zero and will not affect the pricing mechanism in any way.}  If $0\leq SDR_t<1$, indicating over-demand in $t$, then the unmet demand bids will purchase energy at a pre-defined utility rate, denoted as $UR$. Without loss of generality, we assume that $FIT < UR$.\footnote{If $FIT > UR$, it means that energy consumers pay more than the utility rate to purchase energy from the prosumers, which is equivalent to a direct subsidy from energy consumers (including low-income consumers who do not have the means to invest in DERs) to prosumers. This cannot be justified from an equity perspective.} 

\begin{figure}[!htb]
    \centering
    \includegraphics[scale = 0.35]{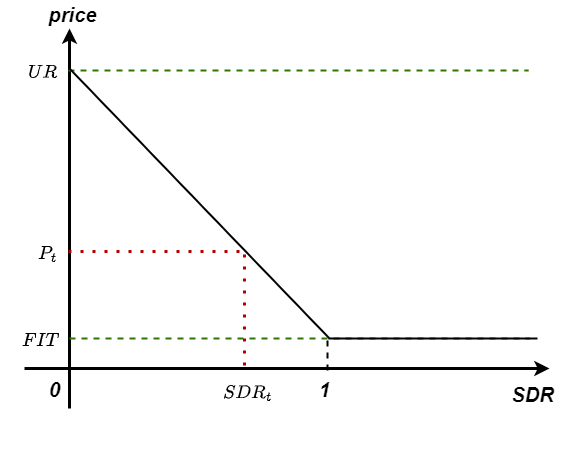}
    \caption{Market clearing using SDR}
    \label{fig:SDR_Price}\vspace*{-5pt}
\end{figure}

The SDR mechanism determines the market clearing price as follows:
\begin{align}
	\label{eq:Price_SDR}
	 P_t:=P(SDR_t) = 
	\begin{cases}
		(FIT-UR)\cdot SDR_t + UR, &  0 \leq SDR_t \leq 1;\\
		FIT, & SDR_t > 1.
	\end{cases}
\end{align}
In \eqref{eq:Price_SDR}, $P_t$ denotes the market clearing price in round $t$, which is a piece-wise linear function with respect to the $SDR_t$, as illustrated in Figure \ref{fig:SDR_Price}. 

While the SDR approach may be criticized for not being based on economic theories, we argue that its simplicity and transparency make it well-suited for a P2P energy market with zero marginal cost resources. In our previous work \cite{zhao2021auction}, we identified that a P2P energy market using a traditional double-sided auction is unlikely to perform well due to its unique characteristics. Specifically, uncleared electricity demand bids must be bought at the $UR$ rate, and uncleared energy supply bids must be sold at $FIT$, with both rates known to all market participants.  The public information of $UR$ and $FIT$ naturally leads sellers to bid at $UR$ and buyers to bid at $FIT$, aiming to maximize their payoffs, since the marginal generation costs are zero. 
This will result in a bang-bang market outcome: when there is overdemand, the market price is $UR$; when there is oversupply, the price is $FIT$. We view the SDR approach as a fair and reasonable alternative to the marginal-cost-based approach to avoid such outcomes, especially in a new market where supply is likely much less than demand. 

Additionally, the SDR method offers two key benefits: First, it maintains the market clearing price between $UR$ and $FIT$, ensuring that transactions for buyers and sellers are at least as favorable as dealing directly with the utility. Second, it simplifies the bidding process by requiring only quantity bids. This allows consumers with flexible loads to focus on adjusting their demand timings, and prosumers to better manage their energy storage and sales strategies, without the complexity of submitting price-quantity pairs.

\begin{figure}[!htb]
    \centering
    \hspace*{-10pt}
    \includegraphics[scale = 0.46]{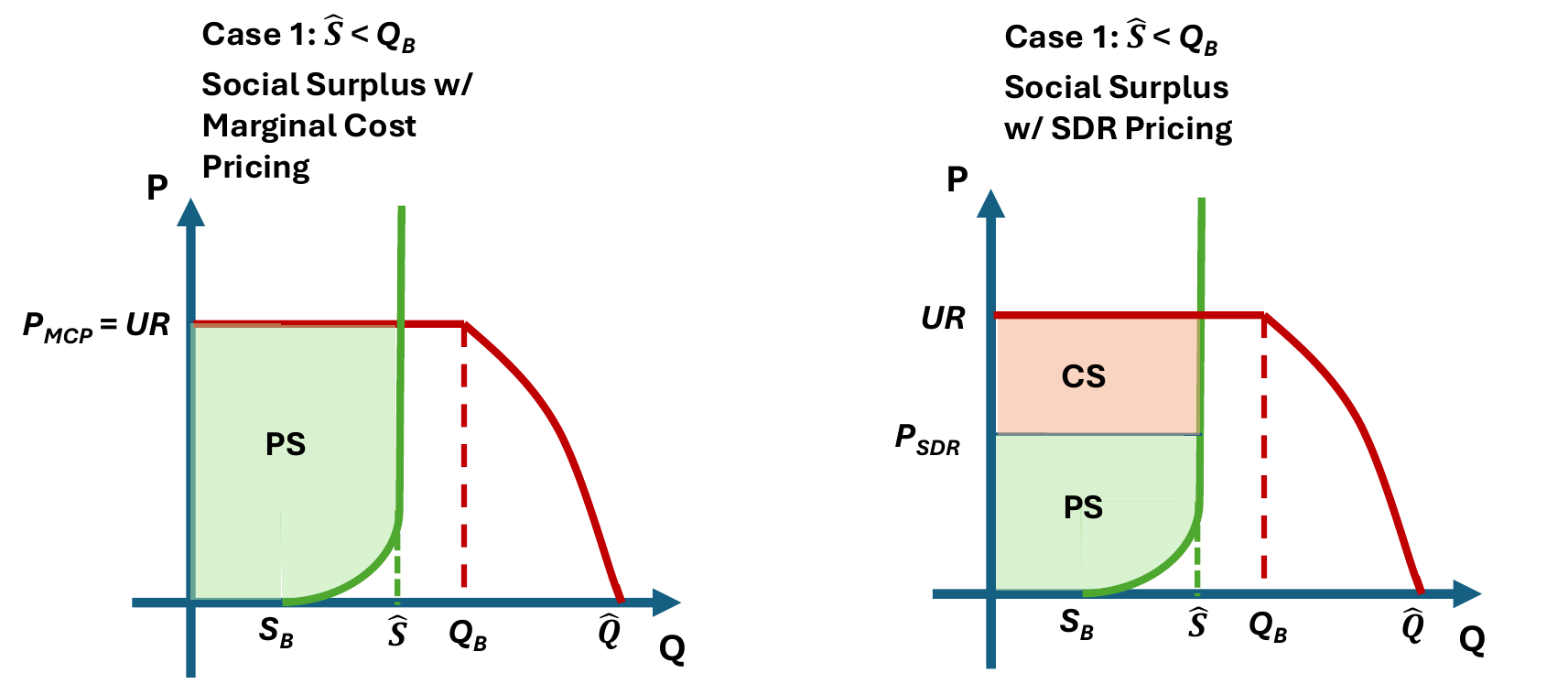}
    \caption{Social surplus comparison: total zero-marginal-cost supply less than total inflexible load}
    \label{fig:SS_Case1}\vspace*{-5pt}
\end{figure}

\begin{figure}[!htb]
    \centering
     \hspace*{15pt}
    \includegraphics[scale = 0.45]{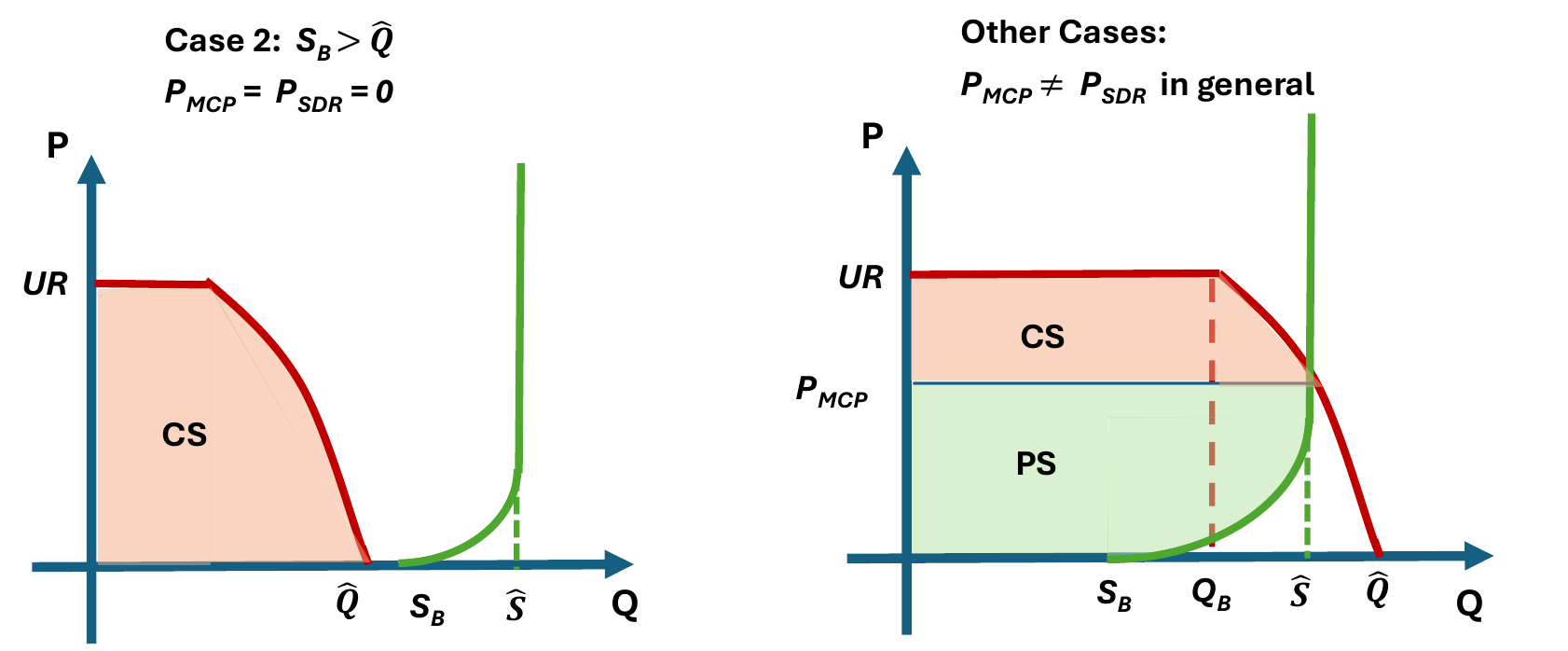}
    \caption{Social surplus: total zero-marginal-cost supply more than total inflexible load }
    \label{fig:SS_Case2}\vspace*{-5pt}
\end{figure}

With the above being said, we are interested in comparing consumer and producer surplus (and their sum, social surplus) between the marginal-cost-pricing (MCP) approach and the SDR approach. Conceptually, consumer surplus is the difference between the maximum amount consumers are willing to pay for a good and the amount they actually pay. Similarly, producer surplus represents the difference between the price producers receive for a good and the cost of producing it. However, several complicating factors arise in surplus analysis here. First, both demand and supply have inflexible and flexible components. For electricity consumption, there is inflexible demand (the so-called baseload) and flexible demand, with flexibility either coming from shiftable loads (like EV charging, washers/dryers) or from charging energy storage. For baseload, we do not know the (aggregated) consumers' true willingness to pay. We follow the approach in \cite{FTR} and use the utility rate as a proxy for consumers' willingness to pay when calculating consumer surplus for the inflexible load; that is, {\it{(UR - clearing price) $\times$ baseload}}. For flexible load, we assume consumers' willingness to pay is represented by a downward-sloping curve, as shown in Figures \ref{fig:SS_Case1} and \ref{fig:SS_Case2}. In these figures, $Q_B$ denotes the total baseload, and $\widehat{Q}$ denotes the total load, both flexible and inflexible. For supply, we have inflexible supply, such as grid-tied solar (that is, there is no storage). In this case, any energy generated beyond the prosumer's own consumption is automatically sold back to the grid. We denote the total inflexible supply by $S_B$. There is also flexible supply, where flexibility comes entirely from the option to store energy for later sale. For flexible supply, we assume prosumers' willingness to sell is represented by an upward-sloping curve until it hits the total supply capacity (denoted by $\widehat{S}$), as depicted in Figures \ref{fig:SS_Case1} and \ref{fig:SS_Case2}.

Next, we analyze the differences in surpluses between MCP and SDR. We assume that under MCP, consumers and prosumers bid their true willingness-to-buy and willingness-to-sell curves, respectively. We must consider different cases we do not know how prosumers would bid under the SDR mechanism. The first case is when $\widehat{S} < Q_B$, meaning the maximum prosumers' supply is less than the total baseload. In this scenario, under MCP, the market clearing price will be $UR$, and prosumers capture all the surplus (denoted as PS in the figures), as shown in the left graph of Figure \ref{fig:SS_Case1}. In the SDR case, regardless of the agents' strategies, we will have $0 < SDR < 1$, since the total supply is less than demand. In this case, consumers capture the surplus of $(UR - P_{SDR}) \times \widehat{S}$, as shown in the right graph of Figure \ref{fig:SS_Case1}, with the area denoted by CS, and it is easy to see that the social surplus is identical to the MCP case. This is likely the most common scenario now and in the near future, where prosumers' total solar and storage capacity is less than total inflexible demand. The second case is the opposite extreme, where $S_B > \widehat{Q}$. In this case, $P_{SDR} = 0$, which is the same as $P_{MCP}$. As a result, consumers capture all the surplus, and the social surplus is again identical between MCP and SDR, as shown in the left graph of Figure \ref{fig:SS_Case2}. In cases beyond these two extremes, the situation is less clear, as it depends on how agents bid in both the SDR and MCP markets. In such cases, we expect that $P_{SDR}$ and $P_{MCP}$ will differ, making the analysis of social surplus more complex. We defer this analysis to future research.

It is important to note, based on Figures  \ref{fig:SS_Case1} and \ref{fig:SS_Case2}, that the MCP approach would not present a problem if the willingness-to-buy and willingness-to-sell curves were known, even when all supply resources have zero marginal costs, as market clearing prices would not necessarily be 0 or $UR$. However, determining these curves is impractical, as even consumers and prosumers often do not know their precise willingness to buy or sell in relation to flexible demand and supply. This is exactly where the SDR approach offers a distinct advantage: it removes the need for agents to define their true preferences -- which are inherently uncertain—and simply requires them to bid quantities. Moreover, in at least two cases, there is no loss in social surplus, and in fact, the SDR mechanism results in greater consumer surplus when demand significantly exceeds prosumer supply.

\section{Decentralized Learning  with Consensus Updates}
\label{sec:Consensus}
In this section, we first formulate the decision-making problem faced by each agent as a MDP. (For brevity, we only describe a prosumer's problem. A pure consumer's problem is similar, just without the supply bids.) 
 We then introduce the details of the consensus-MARL algorithm.

\subsection{Markov Decision Process for a Prosumer}
\label{subsec:CMDP}
In our setup, we assume that each prosumer 
locates at one bus of the distribution
network and has two resources under their control: a solar photovoltaic (PV) system with a smart inverter, and a rechargeable battery as energy storage. Each prosumer also has a fixed baseload of energy consumption. At a particular time
period t, whether a prosumer is a net energy seller or a buyer depends on their baseload versus the level of PV generation and energy storage.

Due to the presence of energy storage, each agent’s bidding and charging decisions are linked over time, which are naturally modeled through dynamic programming. Since an agent's decisions at $t$ only depend on its energy storage level at $t-1$ and the PV generation in $t$, a discrete-time MDP model is suitable here. In the following, we describe the key building blocks of each agent's MDP model.  

 \textit{Observation and state variables:} Agent $i \in \mathcal{I}$ at time step $t$ is assumed to receive the system states $s_t = (s_{1,t},\dots,s_{I,t}) \in\mathcal{S}:= \Pi_{i=1}^I \mathcal{S}_i$, which is the concatenation of the states of each agent, where $I$ is the total number of agents. Specifically, the individual state variable is defined as $s_{i,t} := (d^p_{i,t}, d^q_{i,t}, e_{i,t},PV_{i,t}) \in \mathcal{S}_i$, where $d^p_{i,t}$ and $d^q_{i,t}$ are the inflexible demand of active and reactive power of agent $i$, $e_{i,t}$ is the state of charge of agent $i$'s energy storage system, and $PV_{i,t}$ is the PV (real) power generation in $t$. 
 
 Among the state variables, $(d^p_{i,t}, d^q_{i,t}, PV_{i,t})$ are only observations because their transitions adhere to underlying distributions that are independent of agents' actions. In contrast, the transition of $e_{i,t}$ is influenced by agents' actions, with details to be elaborated further below after agents' actions are defined.

 The assumption that each agent is required to observe the whole system's state may appear to be strong in a decentralized setting and is prone to criticisms and privacy concerns. Our responses are twofold: first, the identity of the agents is irrelevant, and hence, from an agent's perspective, which state variable corresponds to which specific agent is not known. 
 Second,
 the proposed algorithms are intended to be implemented on devices that can receive information from the grid (generally referring to a utility or a distributed system operator) and automatically participate in the repeated P2P market without human intervention (aka control automation). The information received by the devices should be encrypted and can only be decrypted by the control-automation devices. Granted that the devices could be compromised, there have been works on how to address adversarial attacks in a multi-agent reinforcement learning framework \cite{Gupta}, which leads to interesting future research directions. 

\textit{Action or control variables:} at time $t$, the actions that agent $i$ can take are represented by a vector  $a_{i,t} := (a^{q}_{i,t}, a^e_{i,t}) \in \mathcal{A}_i = \mathcal{A}^q_i \times \mathcal{A}^e_i$, where $a^{q}_{i,t}$ is the reactive power injected or absorbed by agent $i$'s smart inverter, and $a^{e}_{i,t}$ is energy charged (if $a^{e}_{i,t} > 0$) to or discharged (if $a^{e}_{i,t} < 0$) from the battery. 
The feasible action spaces of $a^{q}_{i,t}$ and $a^e_{i,t}$ are generically denoted by 
$\mathcal{A}^q_i$ and $\mathcal{A}^b_i$, respectively.
In the specific context of our model, it is safe to assume that the action space  $\mathcal{A}_i = \mathcal{A}^q_i \times \mathcal{A}^e_i $ is bounded. 

\textit{State transition:}
Out of the state variables defined earlier, the energy storage level $e_{i,t}$ is the only one that is directly affected by an agent's own action. The state transition for $e_{i,t}$ can be written as follows: 
  \begin{align}
  \label{eq:trans}
    &  e_{i,t+1} =  \max \Big\{ \min \Big[e_{i,t} + \eta^c_{i} \max(a^e_{i,t},0)  + \frac{1}{\eta^d_{i}} \min(a^e_{i,t},0), \overline{e}_i\Big], 0\Big\},
  \end{align}
  where $\eta^c_{i}$ and $\eta^d_{i}$ are the charging and discharging efficiency of agent $i$'s battery, respectively, and $\overline{e}_i$ is the battery capacity. The charging efficiency represents the ratio of the amount of energy effectively stored in the system to the energy input during the charging process; while the discharging efficiency is the ratio of the energy delivered by the system during discharge to the energy that was initially stored in it. Their product,  $\eta^c_{i} \times \eta^d_{i}$, yields the so-called round-trip efficiency. 
  The outside ``max" operation in \eqref{eq:trans} is to ensure that the energy level in the battery will not be negative; while the first ``min" operation in the bracket is to ensure that the battery storage capacity is not exceeded. 

    \textit{Reward:} 
   Each agent $i$'s reward function is affected by both the collective actions of the agents and the system states. Specifically, agent $i$'s reward in time step $t$, denoted by $r_{i,t}$, is a random variable whose conditional expectation has three components as follows:
     \begin{align}
     \begin{split}
              \mathbb{E}[r_{i,t}|s_t, a_t] := 
        &\ \mathbb{E}[r^{m}_{i}(a^e_{i,t}; a^e_{-i,t}, s_t)] + \mathbb{E}[r^{v}(a_{i,t}; a_{-i,t}, s_t)]/I \\[5pt]
      &+ \mathbb{E}[r^{l}(a_{i,t}; a_{-i,t}, s_t)]/I.
     \end{split}
      \label{eq:reward}
  \end{align}
  In \eqref{eq:reward}, the notation $a_{-i,t}$ refers to the collection of all other agents' actions, excluding agent $i$'s; while $a^e_{-i,t}$ refers to only the energy charge/discharge decisions of the other agents. 
  The first term $r^{m}_{i}$ in \eqref{eq:reward} is the energy purchase cost or sales profit of agent $i$ from the P2P energy market (with the superscript `m' standing for `market'), whose formulation is as follows: 
  \begin{align}
\label{eq:P2PReward}
	& r^{m}_{i} :=
	\begin{cases}
		\mathbb{I}_{i\in\mathcal{B}_t} \times \Big[ SDR_t \cdot P_t \cdot b_{i,t}  
		+\  (1-SDR_t) \cdot UR \cdot b_{i,t} \Big]\\
		+\ \mathbb{I}_{i\in\mathcal{S}_t} \times \Big(P_t \cdot b_{i,t}\Big) ,\quad \mathrm{if}\ 0 \leq SDR_t \leq 1,\\[10pt]
		FIT\cdot b_{i,t}, \quad \mathrm{if}\ SDR_t > 1,
	\end{cases}
\end{align}
where $\mathbb{I}_{i\in\mathcal{B}_t}$ is an indicator function such that $\mathbb{I}_{i\in\mathcal{B}_t} = 1$ if $i\in \mathcal{B}_t$, and 0 otherwise. Similarly,  $\mathbb{I}_{i\in\mathcal{S}_t} = 1$ if $\in \mathcal{S}_t$ and 0 otherwise. $SDR_t$ is the supply-demand-ratio as defined in \eqref{eq:SDR}, which depends on all agents' bids, and the bids further depend on each agent's charging/discharging decisions, as well as the PV generation and the baseload. Specifically, the bids can be explicitly written as  
\begin{equation} 
\label{eq:action2bid}
b_{i,t} = 
\begin{cases}
    PV_{i,t} - d^p_{i,t} - \min ( a^{e}_{i,t},\frac{\overline{e}_i - e_{i,t}}{\eta^c_{i}}), \ \ &\text{if } a^{e}_{i,t} \geq 0,\\
    PV_{i,t} - d^p_{i,t} - \max ( a^{e}_{i,t}, -e_{i,t}\cdot \eta^d_{i}), \ \ &\text{otherwise},
\end{cases}
\end{equation}
which is simply the net energy of PV generation minus baseload demand (of real power) and charge to the battery. Since PV generation is inherently variable, this net available energy is also a random variable.
The market clearing price $P_t$ in \eqref{eq:P2PReward} is defined in \eqref{eq:Price_SDR}. 

Note that when $SDR_t < 1$, not all demand bids will be cleared in the P2P market at time $t$. In such instances, they will need to purchase the needed energy at the utility rate, leading to the $UR \cdot b_{i,t}$ term when defining the reward $R_i^m$ in equation \eqref{eq:P2PReward}. To ensure fairness and prevent any inadvertent favoritism among the demand bids, we employ a mechanism where each demand bid is partially cleared. Specifically, the clearance proportion is exactly $SDR_t$ (which represents the percentage of total demand bids cleared), and the uncleared bid for every agent is then $(1 - SDR_t) \cdot b_{i,t}$. This approach of proportional scaling echoes the rules used in multi-unit double auctions in \cite{huang2002design}.
  
  The second term in the reward function \eqref{eq:reward}, $r^{v}$, is the system voltage violation penalty at a given time $t$. To provide an explicit form of $r^{v}$, we first write out the standard bus-injection model. Assume that the network has $N$ buses, we have that for $\kappa = 1, \ldots, N$ (the time index $t$ is omitted below for simplicity):
  \begin{align}
  \begin{split}
      p_{\kappa}=&\sum_{j=1}^{N}|V_{\kappa}||V_{j}|(G_{\kappa j} \cos (\alpha_{\kappa}-\alpha_{j}) +B_{\kappa j} \sin (\alpha_{\kappa}-\alpha_{j})), \\
q_{\kappa}=&\sum_{j=1}^{N}|V_{\kappa}||V_{j}|(G_{\kappa j} \sin (\alpha_{\kappa}-\alpha_{j}) -B_{\kappa j} \cos (\alpha_{\kappa}-\alpha_{j})).
  \end{split}
  \label{eq:BusInjection2}
\end{align}
In the above system of nonlinear equations, $p_{\kappa}$'s and $q_{\kappa}$'s are input data, with $p_{\kappa}$ being the net real power injection (or withdrawal) at bus $\kappa$ (i.e. $p_{\kappa} = \sum_{i\in \mathcal{N}_{\kappa}} b_{i}$, with $b_i$ given in \eqref{eq:action2bid} and $\mathcal{N}_{\kappa}$ denoting the set of agents at bus $\kappa$), 
and $q_k$ being the net reactive power flow at bus $k$ (i.e. $q_{\kappa} = \sum_{i\in \mathcal{N}_{\kappa}} (- d^q_i + a^q_i)$). The variables are $|V_{\kappa}|$ and $\alpha_{\kappa}$ for $\kappa = 1, \ldots, N$, with $|V_{\kappa}|$ being the voltage magnitude and $\alpha_{\kappa}$ the phase angle at bus $\kappa$. 
The terms $G_{\kappa j}$ and $B_{\kappa j}$ are parameters that represent the real and imaginary parts of the admittance of branch $\kappa{\text -}j$ in the network, respectively. 

At a given round $t$ of the P2P trading, for a given set of the real and reactive power injection/withdrawal as the result of agents' bidding, if the system of equations \eqref{eq:BusInjection2} has a solution,\footnote{Sufficient conditions under which \eqref{eq:BusInjection2} has a solution are provided in \cite{Sun}.} then we can define the voltage violation penalty as follows:
  \vspace*{-5pt}
  \begin{align}
      r^{v}:= - \lambda \sum_{\kappa=1}^N \text{clip} \Big[\max (|V_{\kappa,t}|-\overline{V}_{\kappa}, \underline{V}_{\kappa}-|V_{\kappa,t}|), \{0, M\}\Big].  
 \label{eq:penalty}   \vspace*{-5pt}
  \end{align}  
  In \eqref{eq:penalty}, $\overline{V}_{\kappa}$ and $\underline{V}_{\kappa}$ represent the upper and lower limit of bus $k$'s voltage magnitude, respectively. 
 Let $f_{clip}$ denote a generic `clip' function in the form of $f_{clip} = \text{clip}[x, \{v_{min}, v_{max}\}]$, which is defined as $f_{clip} = x$ when $v_{min} \leq x \leq v_{max}$, $f_{clip} = v_{min}$ if $x <  v_{min}$, and 
 $f_{clip} = v_{max}$ if $x >  v_{max}$. 
 Based on the clip function definition, if the voltage magnitudes $|V_{\kappa,t}|$ obtained from solving equations \eqref{eq:BusInjection2} are within the voltage limit of $[\underline{V}_{\kappa}, \overline{V}_{\kappa}]$ on every bus $\kappa = 1, \ldots, N$, then $R^v = 0$; if there is voltage violation at a certain bus $\kappa$, 
 the term $\max (|V_{\kappa,t}|-\overline{V}_{\kappa},\  \underline{V}_{\kappa}-|V_{\kappa,t}|)$ becomes positive, and $r^v$ becomes a negative number with an arbitrary positive number $\lambda$ to amplify the voltage violation. The other positive parameter in \eqref{eq:penalty}, $M$, is just to ensure that the reward $r^v$ is bounded below, a technical condition needed to ensure the convergence of the consensus MARL algorithm to be introduced in the next section. If the power flow equations \eqref{eq:BusInjection2} do not have a solution with a given set of agents' bids, we can simply set $ r^{v} = - \lambda \cdot N \cdot M$. 
 
 The third term in the reward function \eqref{eq:reward}, $r^{l}$, is the line capacity violation penalty over the entire distribution network at a given time $t$. Let $S_{\kappa,j}$ denote the apparent power flow from bus $\kappa$ to bus $j$, which is defined as $ S_{\kappa,j} = V_j I_{\kappa,j}^*$, 
where $I_{\kappa,j}^*$ is the complex conjugate of $I_{\kappa,j}$ -- the current from bus $\kappa$ to bus $j$.  $I_{\kappa,j}$ is calculated $
I_{\kappa,j} = Y_{\kappa,j}(V_{\kappa} - V_j)$, 
where $Y_{\kappa,j}$ is the complex admittance of the branch between buses $\kappa$ and $j$, and $V_{\kappa}$ and $V_j$ are the voltages from the power flow equations in \eqref{eq:BusInjection2}. Let $\mathcal{E}$ be the set of branches with constraints on the magnitude of the apparent power flow. The total line capacity violation penalty can then be defined as follows:
  \vspace*{-5pt}
  \begin{align}
      r^{l}:= - \gamma \sum_{(\kappa,j)\in\mathcal{E}} \min \Big[\max (|S_{\kappa,j}|-\overline{S}_{\kappa,j}, 0),C\Big],
 \label{eq:c_penalty}   \vspace*{-5pt}
  \end{align}  
where $\overline{S}_{\kappa,j}$ represent the upper limit of branch $\kappa-j$, $C$ is a constant to ensure that the reward $r^l$ is bounded below, and $\gamma$ is an arbitrary positive number to amplify the capacity violation. If \eqref{eq:BusInjection2} does not have a solution, similar to the voltage violation case, we set $ r^{l} = - \lambda \cdot N \cdot C$. 

We want to clarify that the voltage and line capacity violation penalties are not actual financial penalties for the agents, as market clearing occurs before real-time energy delivery. Instead, these penalties only serve as feedback to help agents refine their reinforcement learning policies. Note that the (fictitious) network violation penalties should not be confused with the fees utilities charge for grid maintenance. This trading market does not account for how utilities impose transmission and distribution (T\&D) maintenance fees, which are typically fixed rather than volumetric in most U.S. regions \cite{OptimalTau}. We can adjust the utility rate (UR, the price ceiling in Fig. \ref{fig:SDR_Price}) to subtract the fixed charge, and the framework in this paper remains applicable, with the understanding that each transaction will have to pay an additional grid maintenance fee.

After market-clearing and solving for voltage magnitudes, the total reward for agent $i$ at time step $t$ is realized, as defined in \eqref{eq:reward}. Note that while we do not explicitly assume to have a DSO in the distribution network, an entity is still needed to solve Equation \eqref{eq:BusInjection2}. A utility company can assume such a role. The solution process may even be carried out by a distributed ledger system, such as on a blockchain.

 \textit{Objective function:} At each $t$, agent $i$ receives the system state $s_t$,  chooeses an action $a_{i,t} \in \mathcal{A}_i$, and receives a reward $r_{i,t}$. Agent \( i \)'s decision, given the system state \( s_t \in \mathcal{S}\), is determined by the policy \( \pi_{\theta_i}(\cdot|s_t) \). This policy is a density function mapping from \( \mathcal{A}_i \) to [0, $\infty$) and is parameterized by \( \theta_i \). Here, \( \theta_i \) is an element of a generic set \( \Theta_i \in  \mathcal{R}^{m_i} \), where the dimensionality $m_i$ is determined by each agent. In choosing $m_i$, 
each agent faces a tradeoff: it should be large enough to avoid underfitting, yet not too large to risk overfitting and increased training time.
Let $\theta := [\theta_1, \cdots, \theta_I]$ and $\pi_{\theta}(\cdot|s) = \prod_{i=1}^I \pi_{\theta_i}(\cdot | s)$ be the joint policy of all agents. In the following discussion of this section, we assume that the Markov chains $\left\{s_t\right\}_{t \geq 0}$ induced by the collective policy $\pi_{\theta}$ is ergodic with a stationary distribution $\rho_{\theta}$.

Different from a fully decentralized framework, the goal for each agent in the consensus-update framework is to solve the following optimization problem:
  \begin{align}
  \label{eq:payoff}
      \sup_{\theta\in \Theta:= \Pi_{i=1}^I \Theta_i} J(\theta) = \mathbb{E}_{s_0 \sim \rho_{\theta}}\Bigg[\lim_{T\to \infty}\frac{1}{T}\sum_{t=0}^{T} \Bigg(\frac{1}{I} \sum_{i=1}^I r_{i,t}\Bigg)\Bigg|s_0, \pi_{\theta} \Bigg], 
  \end{align} 
  which is to optimize the average system long-term reward. The objective function resembles a cooperative game, which may face similar critiques regarding the need for each agent to access system-wide state variables in a decentralized setting. A pertinent response, as previously mentioned, is the use of control automation: the objective function can be integrated into intelligent control devices, making them `black boxes' for end-users. Additionally, in developing a MARL framework with provable convergence, some level of global information sharing among agents is likely unaoidable. The collective objective function, as outlined in \eqref{eq:payoff}, is essential for decentralized agents to reach a consensus.
  
  To ensure that the objective function in \eqref{eq:payoff} is well-defined, we have the following result. 
  \begin{lemma} The supreumum in \eqref{eq:payoff}, $\sup_{\theta} J(\theta)$, exists and is finite.
      \label{lemma1}
  \end{lemma}
  \begin{proof}
      Since the market clearing price $P_t$, the supply-demand ratio $SDR_t$, and agents bid $b_{i,t}$ for $i = 1, \ldots, I$ are all bounded above, the first term in an agent's reward function,  $R^{m}_{i}$, is then bounded above based on Eq. (\ref{eq:P2PReward}), for any $a \in \mathcal{A}:= \Pi_{i=1}^I \mathcal{A}_i$ and $s\in \mathcal{S}$. 
      The second term,  $R^{v}$, is bounded above by 0 based on its definition in Eq. (\ref{eq:penalty}). Hence, agent $i$'s reward $r_{i,t}$ is bounded at any $t$, and for any $a \in \mathcal{A}$ and $s\in \mathcal{S}$. Since the number of agents is finite, there exists a common upper bound for all $r_{i,t}$'s, and let us denote it as $\overline{R}$. Then by the formulation in \eqref{eq:payoff}, $J(\theta) \leq  \overline{R}$ for all $\theta \in \Theta$. By the well-known least-upper-bound property of real numbers, we know that $\sup_{\theta} J(\theta)$ exists and is finite.
  \end{proof}

\subsection{Consensus-update Actor-critic Algroithm}
\label{subsec:alg}
In this section, we introduce the consensus-update actor-critic algorithm for MARL with continuous state and action spaces, developed in \cite{zhang2018networked}, and apply it to solve Problem \eqref{eq:payoff}. First, define the global relative action-value function $Q_{\theta}(s,a)$ for a given $a\in \mathcal{A}$,  $s\in \mathcal{S}$ and a policy $\pi_{\theta}$ as:
\begin{align}
    Q_\theta(s, a) :=\sum_{t=0}^{\infty} \mathbb{E}\left[\frac{1}{I} \sum_{i=1}^I r_{i,t}-J(\theta) \mid s_0=s, a_0=a, \pi_{\theta}\right]. 
\end{align}
 Note that since all the agents share the same $J(\theta)$, the $Q$ function does not need to have an agent index and is the same across the agents. 

 The function $Q_{\theta}(s,a)$ cannot be calculated by each agent, even if the global state and action information are shared, since the joint policy distribution $\pi_{\theta}$ is not known. Instead, each agent uses $\hat{Q}(\cdot,\cdot;\omega^i):\mathcal{S}\times\mathcal{A}\to\mathcal{R}$, which is a class of functions parametrized by $\omega^i \in \mathcal{R}^K$, where $K$ is the dimension of the $\omega^i$,   to approximate the action-value function $Q_{\theta}(s,a)$. Note that unlike the parameters $\theta_i$, where agents can choose their own policy models (and hence different dimensions of $\theta_i$), the dimension of $\omega^i$ has to be the same across all agents (and hence $K$ does not have an agent index) to facilitate the consensus update.

\begin{figure*}[!htb]
    \centering
    \includegraphics[width=\textwidth]{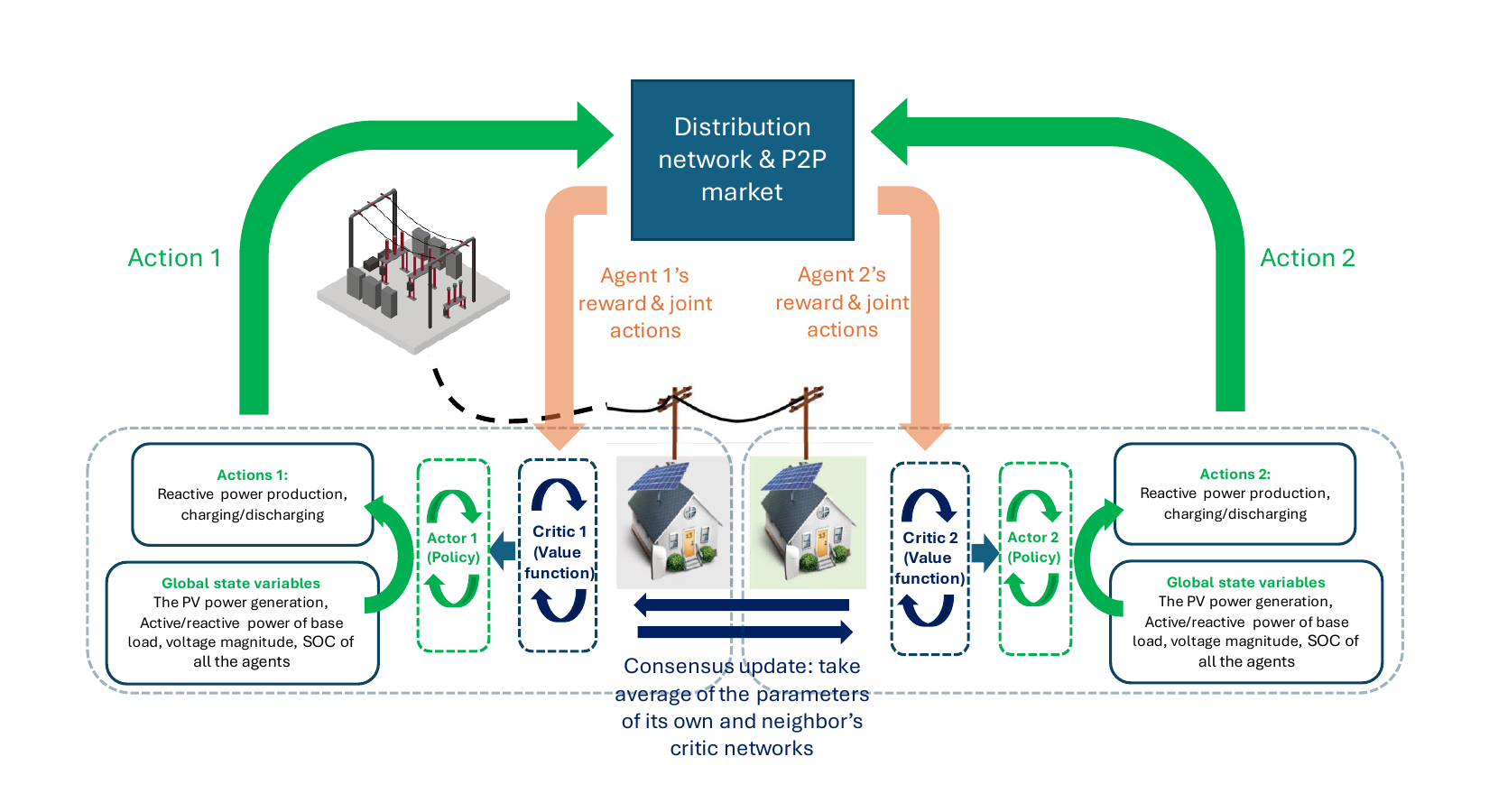}
    \caption{Consensus MARL Framework for the Voltage Control with P2P Market}
    \label{fig:consensus}\vspace*{-5pt}
\end{figure*}

The decentralized consensus-update actor-critic algorithm for networked multi-agents works as follows: at the beginning of time step $t+1$, each agent $i$ observes the global state $s_{t+1}$ and chooses their action $a_{i,t+1}$ according to their own policy $\pi_{\theta_{i,t}}$ (referred to as the `actor'), where $\theta_{i,t}$ is the policy parameter for agent $i$ at time step $t$. Then each agent $i$ will receive the joint action $a_{t+1} = (a_{1,t+1},\dots, a_{I,t+1})$ and their own reward $r_{i,t+1}$, which help them update $\hat{Q}(\cdot,\cdot;\omega^i_t)$ (referred to as the `critic') and $\pi_{\theta_{i,t}}$ on their own in each time step.  
Specifically, each agent first updates the temporal-difference (TD) error $\delta^i_t$ and long-term return estimate $\Bar{r}^i_{t+1}$ as follows: 
\begin{align}
    &\delta_t^i := r_{i,t+1}-\Bar{r}^i_{t+1}+\hat{Q}\left(s_{t+1},a_{t+1};\omega_t^i\right)-\hat{Q}\left(s_{t},a_{t};\omega_t^i\right)\\
    &\Bar{r}^i_{t+1} \leftarrow (1-\beta_{\omega,t}) \cdot \Bar{r}^i_{t} + \beta_{\omega,t}\cdot r_{i,t}
\end{align}
Let $\widetilde{\omega}_t^i$ denote a temporary local parameter of $\hat{Q}(\cdot,\cdot;\omega^i_t)$ for agent $i$, and it is updated as: 
\begin{align}
    \widetilde{\omega}_t^i = \omega_t^i+\beta_{\omega, t} \cdot \delta_t^i \cdot \nabla_\omega \hat{Q}\left(s_{t},a_{t};\omega_t^i\right). \label{eq:tilde_omega}
\end{align}
The local parameter of the policy $\pi_{\theta_{i,t}}$ is updated as
\begin{align}
    \theta_{i,t+1} = \Gamma^i[\theta_{i,t}+\beta_{\theta, t} \cdot \widehat{I}_t^{Q, i}],
    \label{eq:theta_update}
\end{align}
where $\Gamma^i: \mathbb{R}^{m_i} \rightarrow \Theta_i \subset \mathbb{R}^{m_i}$ is a projector operator. An example is the orthogonal projector mapping: $\Gamma^i(x) \triangleq \underset{y \in \Theta_i}{\arg \min }\|y-x\|, \ \forall \mathbf{x} \in \mathbb{R}^{m_i} $, where $\Theta$ is a compact convex set.  The $\widehat{I}_t^{Q, i}$ term in \eqref{eq:theta_update} is given as follows: 
\begin{align}
\begin{split}
    & \widehat{I}_t^{Q, i}=\\
& \int_{\mathcal{A}^i} d \pi_{\theta_{i,t}}\left(a^i \mid s_t\right) \nabla_{\theta_i} \log \pi_{\theta_{i,t}}\left(a^i \mid s_t\right) \hat{Q}\left(s_t, a^i, a_t^{-i} ; \omega_t^i\right).
\end{split}
\end{align}
To limit communication among agents, each agent $i$ only shares local parameter $\widetilde{\omega}_t^i$ with each other; no other information needs to be shared. A consensual estimate of $Q_{\theta}(s,a)$ can then be estimated through updating the parameter $\omega_{t+1}^i$ as follows: 
\vspace*{-6pt}
\begin{align}
    \omega_{t+1}^i = \sum_{j=1}^I \widetilde{\omega}_t^j.
\end{align}
To apply such an algorithm to P2P energy trading, we illustrate the conceptual framework in Fig. \ref{fig:consensus}. The detailed algorithm is presented in  Algorithm \ref{algo1}. 

\begin{algorithm}
\caption{Consensus-update actor-critic algorithm for voltage control with P2P energy market}
\textbf{Input:} Initial values of the parameters $\Bar{r}_0^i, \omega_0^i, \widetilde{\omega}_0^i, \theta_0^i, \forall i \in \{1,\dots,I\}$, the initial state $s_0$, and stepsizes $\left\{\beta_{\omega, t}\right\}_{t \geq 0} \text { and }\left\{\beta_{\theta, t}\right\}_{t \geq 0}$.\\
\For{\textbf{each} agent $i = 1,...,I$}
{
Makes the decision on reactive power generation and battery charging/discharging $a_{i,0} \sim \pi_{\theta_0^i}\left(\cdot \mid s_0\right)$.
}The power flow equation \eqref{eq:BusInjection2} is solved, the P2P energy market is cleared, and the information of joint actions $a_0=\left(a_{i,0}, \ldots, a_{I,0}\right)$ is sent to each agent.\\
\For{\textbf{each} agent $i = 1,...,I$}
{
Observes the reward $r_{i,0}$.
}
\For{\textbf{each} time step $t = 0,1,...T$}
{
\For{\textbf{each} agent $i = 1,...,I$}
{
Observes the next global state $s_{t+1}$.\\
Update $\Bar{r}^i_{t+1} \leftarrow (1-\beta_{\omega,t}) \cdot \Bar{r}^i_{t} + \beta_{\omega,t}\cdot r_{i,t}$.
Makes the decision on reactive power generation and battery charging/discharging $a_{i,t+1} \sim \pi_{\theta_{i,t}}\left(\cdot \mid s_{t+1}\right)$.
}
The power flow equation \eqref{eq:BusInjection2} is solved, the P2P energy market is cleared, and the information of joint actions $a_{t+1}=\left(a_{i,t+1}, \ldots, a_{I,t+1}\right)$ is sent to each agent.\\
\For{\textbf{each} agent $i = 1,...,I$}
{
1. Observes the reward $r_{i,t+1}$ and update $\delta_{t}^i \leftarrow r_{i,t+1}-\Bar{r}^i_{t+1}+\hat{Q}\left(s_{t+1},a_{t+1};\omega_t^i\right)-\hat{Q}\left(s_{t},a_{t};\omega_t^i\right)$.\\
2. \textbf{Critic step: } $\widetilde{\omega}_t^i \leftarrow \omega_t^i+\beta_{\omega, t} \cdot \delta_t^i \cdot \nabla_\omega \hat{Q}\left(s_{t},a_{t};\omega_t^i\right)$.\\
3. \textbf{Actor step: } $\theta_{i,t+1} \leftarrow \Tilde{\Gamma}^i[\theta_{i,t}+\beta_{\theta, t} \cdot \widehat{I}_t^{Q, i}]$, where $\Gamma^i: \mathbb{R}^{m_i} \rightarrow \Theta_i \subset \mathbb{R}^{m_i}$, and
\begin{align}
& \widehat{I}_t^{Q, i}=  \int_{\mathcal{A}^i} d \pi_{\theta_{i,t}}\left(a^i \mid s_t\right)  \cdot \noindent\\
&\nabla_{\theta_i} \log \pi_{\theta_{i,t}}\left(a^i \mid s_t\right) \hat{Q}\left(s_t, a^i, a_t^{-i} ; \omega_t^i\right),
\end{align}
4. Send $\widetilde{\omega}_t^i$ to the other agent in the system.
}
\For{\textbf{each} agent $i = 1,...,I$}
{
\textbf{Consensus step: } $\omega_{t+1}^i \leftarrow \frac{1}{I}\sum_{j=1}^I \widetilde{\omega}_t^j$.  
}
}
\label{algo1}
\end{algorithm}


\subsection{Convergence Results}
In this section, we discuss the convergence results of Algorithm \ref{algo1}. For ease of notation, we drop the $t$ index of the parameters $\theta_i$ for $i\in \mathcal{I}$ in the following discussions.  First, we need a technical condition of Markov chains, known as the geometric ergodicity. 

   \begin{definition}(Geometric ergodicity, referred to as uniform ergodicity in \cite{MCMC})
A Markov chain having stationary distribution $\pi(\cdot)$ is geometricly ergodic if 
$$
||P^m(x,\cdot) - \pi(\cdot)|| \leq M c^m,\ m = 1, 2, 3, \ldots
$$
for some $0 < c < 1$ and $0 < M < \infty$, where $P$ is the transition kernel of the Markov chain. 
\end{definition}
The concept of geometric ergodicity specifies the rate at which a Markov chain converges to a stationary distribution if it exists. In the following, we state the first assumption to ensure the consensus MARL convergence to a steady state, which requires the Markov chains formed by the system state and state-action pairs to be both geometrically ergodic. Granted that this assumption cannot be easily verified, it is essential to the convergence proof.

\begin{assumption}
\label{assump1}
    For each $i \in \mathcal{I}, s \in \mathcal{S}$ and $\theta_i \in \Theta_i$, agent $i$'s policy
    is positive; that is, $\pi_{\theta_i}\left(a_i \mid s\right)>0$ for all $a_i \in \mathcal{A}_i$. Additionally, $\pi_{\theta_i}\left(\cdot\mid s\right)$ is assumed to be continuously differentiable with respect to the parameter $\theta_i$ over $\Theta_i$. Finally, the Markov chains $\left\{s_t\right\}_{t \geq 0}$ and $\left\{\left(s_t, a_t\right)\right\}_{t \geq 0}$ induced by the agents' collective policies $\pi_{\theta}$ are both geometrically ergodic, with the coressponding stationary distribution denoted by $\rho_{\theta}$ and $\tilde{\rho_{\theta}}$, respectively.
\end{assumption}
\noindent\textbf{Remark 1} The first part of Assumption \ref{assump1} is a standard assumption on policy functions. One example of such a policy is Gaussian: $\pi_{\theta_i}(\cdot \mid s) \sim$ $\mathcal{N}\left(\eta_{\theta_i}(s), \Sigma_i\right)$,  where $\eta_{\theta_i}(s): \mathcal{S} \rightarrow \mathcal{A}_i \in \mathbb{R}^{n}$ is a fully connected neural network parametrized by $\theta_i$ and is continuously differentiable with respect to $\theta_i$ and $s$, and $\Sigma_i \in \mathbb{R}^{n \times n}$ is the covariance matrix. 

\begin{assumption}
\label{assump2}
The policy parameter $\theta_{i}$ for each agent is updated by a local projection operator, $\Gamma^i: \mathbb{R}^{m_i} \rightarrow \Theta_i \subset \mathbb{R}^{m_i}$, that projects any $\theta_{i}$ onto a compact set $\Theta_i$. In addition, $\Theta=\prod_{i=1}^I \Theta_i$ includes at least one local minimum of $J(\theta)$.
\end{assumption}
\noindent\textbf{Remark 2} The requirement of the local projection operator is standard in convergence analyses of many reinforcement learning algorithms. If $\Theta_i$ is a convex set, a qualifying example of $\Gamma^i$ can be the nearest point projection on $\Theta_i$; that is, 
\begin{align}
    \Gamma^i(\theta_i) = \arg \max_{{\theta_i}^*\in \Theta_i} \Vert \theta_i - {\theta_i}^*\Vert_2.
\end{align}
Next, we make an assumption on the action-value function approximation.
\begin{assumption}
\label{assump3}
For each agent $i$, the action-value function is approximated by linear functions, i.e., $\hat{Q}(s, a ; \omega)=$ $\omega^{\top} \phi(s, a)$ where $$\phi(s, a)=\left[\phi_1(s, a), \cdots, \phi_K(s, a)\right]^{\top} \in \mathbb{R}^K$$ is the feature associated with $(s, a)$. The feature function, $\phi_k: \mathcal{S} \times \mathcal{A} \rightarrow \mathbb{R}$, for $k=1,\cdots,K$, is bounded for any $s \in \mathcal{S}, a \in \mathcal{A}$. Furthermore, the feature functions $\left\{\phi_k\right\}_{k=1}^K$ are linearly independent, and for any $u \in \mathbb{R}^K$ and $u \neq 0, u^{\top} \phi$ is not a constant function over $\mathcal{S} \times \mathcal{A}$.
\end{assumption}
\noindent\textbf{Remark 3.} One exmaple of the action-value function approximation is the Gaussian radial basis function (RBF):
\begin{align}
    \hat{Q}(s, a ; \omega) = \sum_{j=1}^K \omega_i e^{-\gamma_{j}  \Vert [s,a]-c_{j} \Vert } ,
\end{align}
where $[s,a]$ is the concatenation of $s$ and $a$, $\gamma_{j} \in \mathcal{R}^+$ for $j=1,\cdots,K$, $c_{j}\in \mathcal{R}^{|\mathcal{S}|+|\mathcal{A}|}$ for $j=1,\cdots,K$. The parameters $\gamma_{j}$ and $c_{j}$ can be chosen arbitrarily, as long as $(\gamma_{j}, c_{j})$ are different with different $j$ so that the feature functions are linearly independent.
\begin{assumption}
\label{assump4}
The stepsizes $\beta_{\omega, t}$ and $\beta_{\theta, t}$ in Algorithm \ref{algo1} satisfy
\begin{align}
\sum_t \beta_{\omega, t}=\sum_t \beta_{\theta, t}=\infty,\  \mathrm{and} \ \sum_t (\beta_{\omega, t}^2+\beta_{\theta, t}^2)<\infty . 
\end{align}
Also, $\beta_{\theta, t}=o\left(\beta_{\omega, t}\right)$ and $\lim _{t \rightarrow \infty} \beta_{\omega, t+1} \cdot \beta_{\omega, t}^{-1}=1$.
\end{assumption} 
An example of such stepsizes can be $\beta_{\omega, t}= 1/t^{0.8}$ and $\beta_{\theta, t}=1/t$.  

With the above assumptions, the convergence of Algorithm \ref{algo1} is given below. 

\begin{theorem}
\label{Theorem}
Assume that each agent's state space $\mathcal{S}_i$ is compact for $i\in\mathcal{I}$. Under Assumptions 1 -- 4, the sequences $\left\{\theta_{i,t}\right\}$, $\left\{\mu_t^i\right\}$ and $\left\{\omega_t^i\right\}$ generated from Algorithm \ref{algo1} satisfy the following. 

   \begin{itemize}
   \item[(i)] Convergence of the critic step:  $\frac{1}{I}\lim _{t\rightarrow \infty} \sum_{i \in \mathcal{I}} \mu_t^i =J(\theta)$, where $J(\theta)$ is defined in \eqref{eq:payoff}, and $\lim _{t\rightarrow \infty} \omega_t^i=\omega_\theta$ for all $i \in \mathcal{I}$, with $\omega_\theta$ being a solution of a fixed point mapping. (The convergence is in the sense of almost surely.) 
 \item[(ii)] Convergence of the actor step: for each $i\in \mathcal{I}$, $\theta_{i,t}$ converges almost surely to a point $\Hat{\theta}_i$ that is a projection onto the set $\Theta$. 
\end{itemize}
\end{theorem}
\begin{proof} 
The general convergence result of the consensus algorithm is proved in \cite{zhang2018networked} under six assumptions. 
Our Assumption \ref{assump1}, \ref{assump2}, \ref{assump3} and \ref{assump4} directly correspond to four assumptions in \cite{zhang2018networked}. The fifth assumption in \cite{zhang2018networked} regarding agents' time-varying neighborhood is trivially true here since we assume agents can communicate through the whole network, and the network topology does not change over time. Hence, the only remaining item to show is the uniform boundedness of agents' reward $r_{i,t}$. By the definition of the market clearing price $P_t$ in \eqref{eq:Price_SDR}, it is bounded between $FIT$ and $UR$. Together with the compactness assumption of the state space $\mathcal{S}_i$ (which is reasonable since an agent's energy demand and battery/PV capacities are all bounded), the $R_i^m$ component in the reward function, as defined in \eqref{eq:reward}, is uniformly bounded. The other two terms, the voltage violation penalty $R_i^v$, is uniformly upper bounded by 0 and lower bounded by $-\lambda NM$, and the capacity violation penalty $R_i^c$, is uniformly upper bounded by 0 and lower bounded by $-\lambda NC$. Hence, $r_{i,t}$ is uniformly bounded for all $i$ and $t$, and the convergence results follow directly from  Theorem 2 and 3 in \cite{zhang2018networked}.
\end{proof}
\noindent\textbf{Remark 4}.
While the theorem establishes the convergence to a steady state of the critic and actor steps, it does not address how good the system reward is upon convergence. A key concern 
in this context is the potential inadequacy of the linear approximation of the critic function $\hat{Q}(s, a ; \omega)$ in Assumption \ref{assump3}. 
In our model, the distribution network encompasses a multitude of agents, each exhibiting non-linear behaviors and interactions that may not be adequately captured by simple linear models. One natural idea is to deep neural networks (DNNs) instead, as they excel in capturing and modeling non-linear relationships. 
The downside of using DNNs, however, is that the proof of Theorem \ref{Theorem} is no longer applicable. Consequently, the convergence of the critic and actor steps remains an unresolved issue, meriting further investigation in future research.

\noindent\textbf{Remark 5}. To guarantee convergence, all agents must use the consensus algorithm. It has been shown in \cite{Gupta} that this consensus-based MARL algorithm may lack robustness; specifically, a single adversarial attacker could disrupt the joint reward optimization process. A more resilient approach is proposed in the follow-up work \cite{ResilientConcensusMARL}, which could potentially be adapted to enhance the robustness of our framework. While the focus is not necessarily on adversarial agents, it would be valuable to address scenarios where some agents may use different algorithms or act randomly without using any algorithm at all. Exploring such robustness would be a promising future research direction.

Another challenge arises from the dynamic nature of multi-agent environments, where agents frequently enter or leave the system, resulting in an ever-changing environment. The convergence of consensus MARL depends on the exchange of information between neighboring agents (while our results address the exchange of information with all agents in the network, the algorithm is also effective for local neighborhood exchanges, as demonstrated in \cite{zhang2018networked}). As the composition of agents evolves, the joint state and action spaces of neighboring agents can change, potentially impacting the learned policies. Transfer learning \cite{TLSurvey}, which facilitates the transfer of knowledge between domains, may provide a viable solution to adapt policies in such dynamic settings. Specifically, techniques like policy distillation and policy reuse can help preserve useful behaviors while adapting to new conditions. Investigating the application of these techniques in the MARL context offers a promising direction for future research.

\section{Simulation Results}
\label{sec:Sim} 
\subsection{Test Data}
\begin{figure}[!h]
    \centering
    \includegraphics[width=0.6\textwidth]{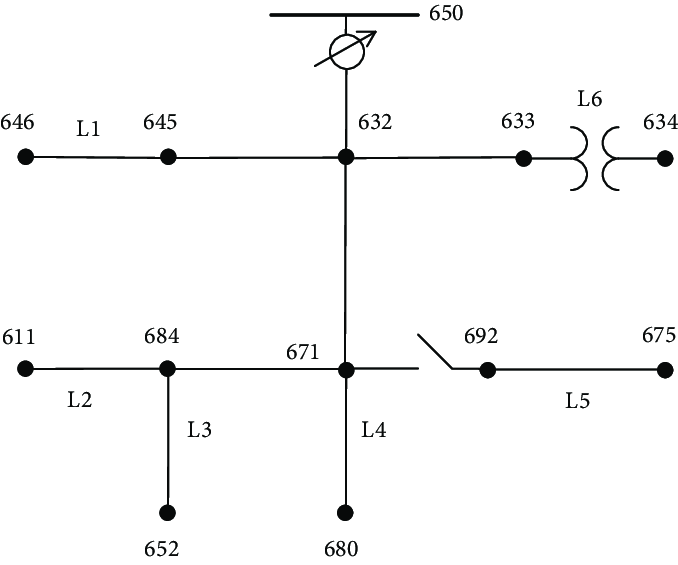}
    \caption{IEEE-13 test feeder}
    \label{fig:feeder}
\end{figure}

We test the algorithm frameworks using the IEEE 13-bus test feeder \cite{13bus}, whose topology is shown in Figure \ref{fig:feeder}. 
 In our simulations, we set each bus to have one prosumer, excluding the substation bus; hence, there are 12 prosumers (and no pure consumers). The length of each time step is set to be 1 hour. The distribution network's voltage limit is set to $[0.98 \ pu,1.01 \ pu]$. The amplifying parameter for voltage violation, $\lambda$, is set at $10^4$. As for the line capacity constraint, we set an apparent power constraint of 15 kVA on the line between Node 684 and 671, and the amplifying parameter for capacity violation, $\gamma$, as $10^2$. The price ceiling and floor, that is, the utility rate $UR$ and feed-in tariff $FIT$, are set at 14 \textcent /kWh and 5 \textcent/kWh, respectively. The capacity of each agent's PV is set to be 20kW, and the energy storage capacity is set to be 50kWh, with the charging and discharging efficiency being 0.95 and 0.9 for each agent. The PV generation of the prosumers and the total base load of the pure consumers on each bus have fixed diurnal shapes. These shapes represent the average values corresponding to each hour of the day. For a detailed explanation and due to page limitations her, please refer to our earlier paper \cite{feng2022decentralized}. Additionally, \cite{feng2022decentralized} provides more information on how the random PV output for each agent is generated.
The power flow equations \eqref{eq:BusInjection2} are solved by the open-source distribution system simulator OpenDSS \cite{EPRI}.

\subsection{Deep Neural Network Implementation}
As stated in Remark 4, we use DNNs instead of linear combinations of features to approximate the critic function $\hat{Q}$. 
To implement the consensus MARL with DNNs, we apply the default architectures of the actor and critic networks of actor-critic algorithms in RLlib \cite{Ray}.  To be specific, for the policy function, we use 
\begin{align}
    \pi_{\theta_i}(\cdot | s) \sim \mathcal{N}\left(\bmat{\sigma^{(1)}_{\theta_i}(s) \\ \sigma^{(2)}_{\theta_i}(s)}, \bmat{\sigma^{(3)}_{\theta_i}(s) & 0\\0 & \sigma^{(4)}_{\theta_i}(s)}\right),
\end{align}
where $\sigma^{(1)}_{\theta_i}(s)$ represents the mean value of the charging/discharging action, and $\sigma^{(2)}_{\theta_i}(s)$ denotes the mean value of the smart inverter action. Similarly, $\sigma^{(3)}_{\theta_i}(s)$ and $\sigma^{(4)}_{\theta_i}(s)$ correspond to the variances of the charging/discharging action and the smart inverter action, respectively. 
The function 
$\sigma_{\theta_i}(s): \mathcal{S} \to \mathcal{R}^4$ is implemented as a DNN, parameterized by $\theta_i$.  This network comprises two fully connected layers, each consisting of 256 neurons, and employs a $tanh$ activate function following each layer. 
A similar neural network architecture is also applied to the approximator of the action-value function $\hat{Q}(\cdot,\cdot;\omega^i)$. The settings of stepsizes are $\beta_{\omega, t}= 1/t^{0.65}$ and $\beta_{\theta, t}=1/t^{0.85}$.

\subsection{Comaprison of Three Different MARL Algorithms}
We compare the consensus MARL with two other algorithm frameworks: (i) a fully decentralized framework in which each agent just solves their own reinforcement learning problem using the PPO algorithm \cite{feng2022decentralized}, while completely ignoring multiagent interactions, and (ii) the MADDPG approach \cite{MADDPG}. 

Different from maximizing the system's total long-term expected reward, the objective for each agent in the fully decentralized framework and in MADDPG is to optimize their own long-term expected reward: 
  \begin{align}
  \label{ownpayoff}
      \sup_{\theta_i} J_i(\theta_i) = \mathbb{E}_{s_0 \sim \rho_{\theta}}\Bigg[\sum_{t=0}^{\infty} \gamma_i^t r_{i,t}|s_0, \pi_{\theta} \Bigg], \ i = 1, \ldots, I,  
  \end{align}  
where $\gamma_i$ is the discount factor of agent $i$. In the fully decentralized framework, each agent uses the single-agent PPO algorithm from \cite{PPO} to update their own policy and value function locally based on only their own state, action, and reward information.

Since all three MARL approaches are categorized as policy gradient descent methods, they update the parameter $\theta_{t+1}$ in a similar manner to approximate gradient ascent of $J(\theta)$ as follows:
\begin{equation}\label{eq:PolicyG}
\theta_{t+1} = \theta_t + \beta \widehat{\nabla J(\theta_t)},
\end{equation}
where $\beta$ is the step size and $\widehat{\nabla J(\theta_t)}$ 
 is a stochastic estimate approximating the gradient of the performance measure. The three MARL approaches differ in how each agent updates $\widehat{\nabla J(\theta_t)}$. To highlight such differences, we present the specific gradient approximations for the three approaches below.

 In the fully decentralized approach, the gradient update uses local states and actions only: for each $i \in \mathcal{I}$, 
\begin{align}
\label{PPO}
    \widehat{\nabla_{\theta_i} J_i\left(\theta_i\right)}=\mathbb{E}_{s \sim \rho^\theta, a_i \sim \pi_{\theta_i}}\left[\nabla_{\theta_i} \log \pi_{\theta_i}\left(a_i \mid s_i\right) Q_i^\pi\left(s_i ; a_i\right)\right].
\end{align}

In the MADDPG approach, each agent makes decisions based on only their own state variables. However, their action-value functions are updated based on the global states and actions of all agents by a central authority, who is assumed to have access to the global information. Specifically, for each agent $i \in \mathcal{I}$,  the gradient approximation in MADDPG is
\begin{align}
\label{MADDPG}
   & \widehat{\nabla_{\theta_i} J_i\left(\theta_i\right)}=  \mathbb{E}_{s \sim \rho^\theta, a_i \sim \pi_{\theta_i}}\left[\nabla_{\theta_i} \log \pi_{\theta_i}\left(a_i \mid s_i\right) Q_i^\pi\left(\bm{s} ; \bm{a_1, \ldots, a_I}\right)\right], 
\end{align}
where $Q_i^\pi\left(\bm{s}; \bm{a_1, \ldots, a_I}\right)$ is the centralized action-value function to be updated by the assumed central authority. 
The policy function, $\pi_{\theta_i}\left(a_i \mid s_i\right) $, on the other hand, uses only local states $s_i$. This is what is referred to as centralized training and decentralized execution. 

For the consensus framework, the gradient estimation uses the so-called expected policy gradient as in \cite{zhang2018networked}:
\begin{align}
\label{CONSENSUS}
    \widehat{\nabla_{\theta_i} J\left(\theta\right)}=\mathbb{E}_{s \sim \rho^\theta, a_{-i} \sim \pi_{\theta_{-i}}} \mathcal{I}_{\theta_i}^Q\left(s, a_{-i}\right),\ i=1, \ldots, I, 
\end{align}
where 
\begin{align}
\label{eq:EPG}
    \mathcal{I}_{\theta_i}^Q\left(s, a_{-i}\right)=\mathbb{E}_{a_i \sim \pi_{\theta_i}} \nabla_{\theta_i} \log \pi_{\theta_i}\left(a_i \mid \bm{s}\right) Q_i^\pi\left(\bm{s} ; \bm{a_1, \ldots, a_I}\right).
\end{align}
While the action-value function $Q_i^\pi$ is the same as in \eqref{MADDPG}, the key difference lies in the policy function. As stated in Section \ref{subsec:CMDP} and reflected in \eqref{eq:EPG}, each agent $i$'s policy depends on all agents' states $\bm{s}$. The other difference between the consensus MARL and the MADDPG algorithm is how the action-value function $Q_i^\pi$ is updated. In MADDPG, the function is updated through centralized training by a central authority; while in consensus MARL, updates are executed via a decentralized consensus process, as described in Section \ref{subsec:alg}.

\subsection{Numerical Results}
\begin{figure}[!htb]
    \centering
    \includegraphics[scale = 0.37]{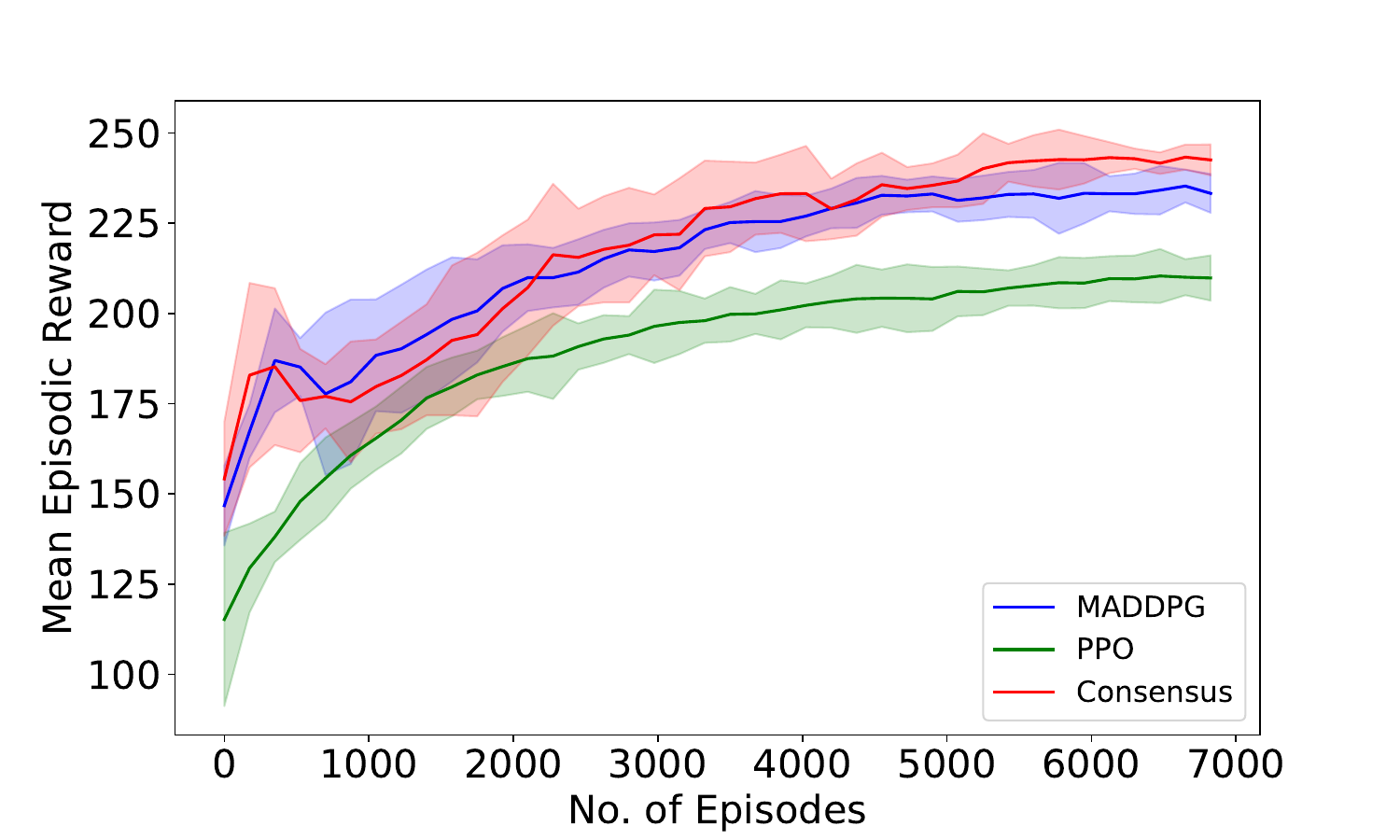}     
    \caption{30-episode moving average of episodic total reward}
    \label{fig:reward_con}\vspace*{-5pt}
\end{figure}

We conduct 20 simulation runs for each of the three frameworks, each comprising 7,000 training episodes, where each training episode is 24 time steps (corresponding to 24 hours in a day). In Figure \ref{fig:reward_con}, the mean of the convergence curves of the 30-episode moving average of the episodic total reward in the distribution network is depicted for the 20 runs of each of the three distinct MARL algorithms. The shaded region represents one standard deviation.
It is evident that the mean episodic total reward gradually converges to a high level at the end of the training in all three frameworks. The results show that the consensus MARL algorithm outperforms the fully decentralized framework, 
which is in line with expectations, since the consensus-update framework benefits from improved approximation of the action-value function through communication. 
A one-sided Welch's t-test yields a p-value of \(7.46 \times 10^{-7}\), strongly suggesting the rejection of the null hypothesis that the episodic total reward of consensus MARL after 7,000 training episodes is less than or equal to that of MADDPG.
 This result provides strong evidence favoring consensus MARL for better performance compared to MADDPG, which is somewhat surprising. Conventionally, one might anticipate a centralized algorithm to outperform decentralized algorithms. The explanation here can be traced to MADDPG’s decentralized execution. In MADDPG, each agent independently updates its actions based on its own local state, despite the centralized nature of the learning process. This approach of relying solely on local information appears to be less effective compared to the consensus update mechanism. 

\begin{figure}[!htb]
    \centering
    \includegraphics[scale = 0.37]{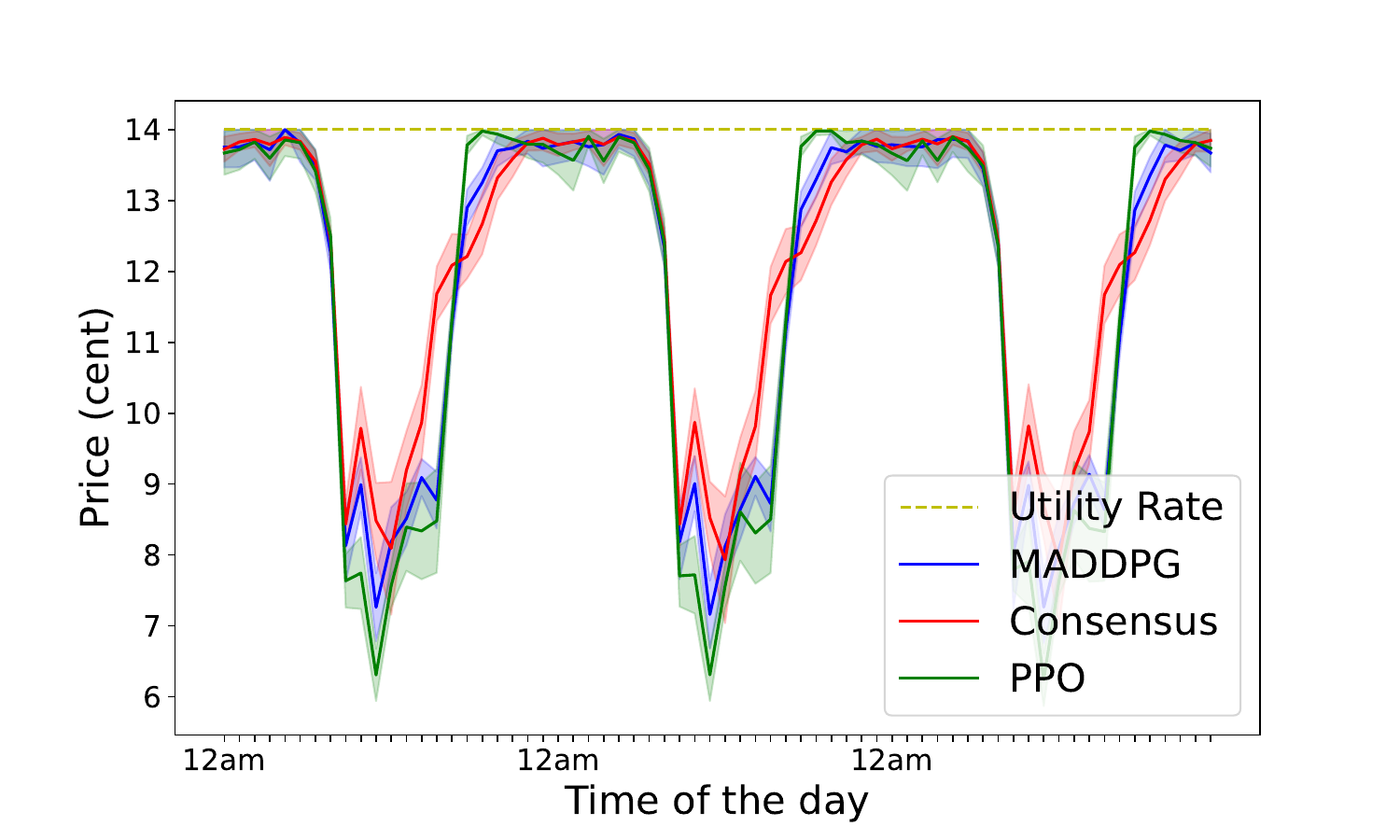}
    \caption{Hourly prices over the last three days}
    \label{fig:price_con}\vspace*{-5pt}
\end{figure}

Figure \ref{fig:price_con} presents the average market prices of the P2P market in chronological order over the last three days after thousands of training episodes.
Notably, in the consensus-update framework, the market price tends to be higher during off-peak hours and lower during peak hours. This suggests that agents using this algorithm have developed more sophisticated trading strategies compared to the other two MARL frameworks. Specifically, the consensus-update algorithm leads to strategies where agents buy more and sell less during off-peak hours, and sell more and buy less during peak hours, unlike the purely decentralized algorithm.

\begin{figure}[!htb]
    \centering
    \includegraphics[scale = 0.37]{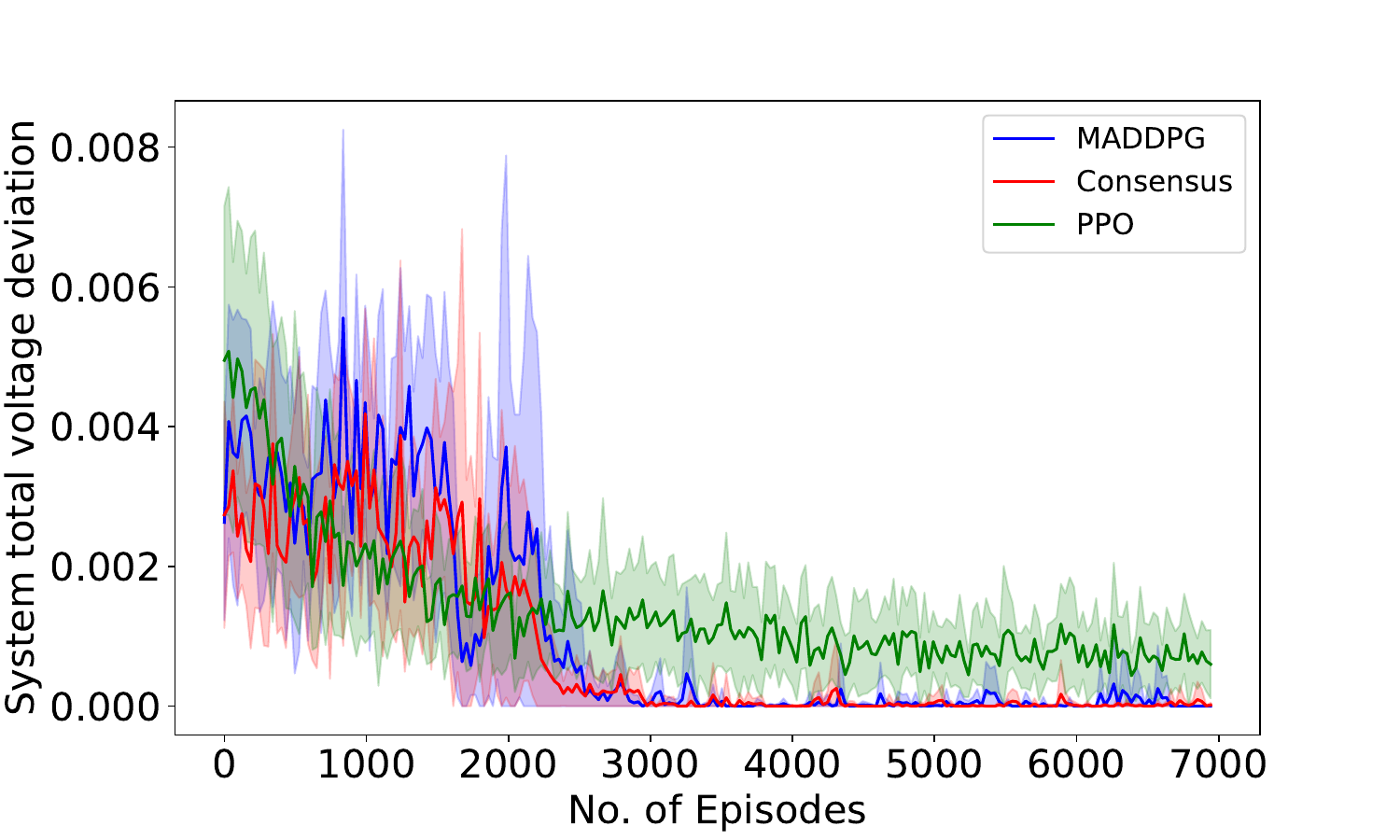}
    \caption{System voltage deviation. The total system voltage deviation (p.u.) in the $n$-th episode is calculated as $\sum_{t=24(n-1)}^{24n-1} \sum_{j:Bus}[\max (0, v^j_{t}-\bar{v})+\max (0, \underline{v}-v^j_{t})]$.}
    \label{fig:violation}
\end{figure}

Figure \ref{fig:violation} shows the gradual reduction of the system voltage deviation after training. 
Initially, it is observed that voltage violations are significantly high across all three frameworks.\footnote{We would like to emphasize that the violations depicted in the figures are not actual violations. They occur in the context of a preliminary stage, such as the hour-ahead market. If agents' bids result in physical network constraint violations, these bids are returned to the agents for them to resubmit till the bids are feasible. In real-time operations, deviations from the cleared bids are inevitable. In such cases, we assume the DSO, typically a utility company, would have other resources available to ensure grid reliability and stability. The costs associated with maintaining safety and stability could be distributed among all market participants, similar to how T\&D  costs are allocated, as mentioned earlier.}  However, the consensus and MADDPG framework demonstrate a rapid decline in these violations, converging swiftly and effectively to zero at the fastest rate at the later stage. The fully decentralized framework demonstrates a slower rate of convergence, and is seemingly unable to fully eliminate voltage violations. 

\begin{figure}[!htb]
    \centering
    \includegraphics[scale = 0.37]{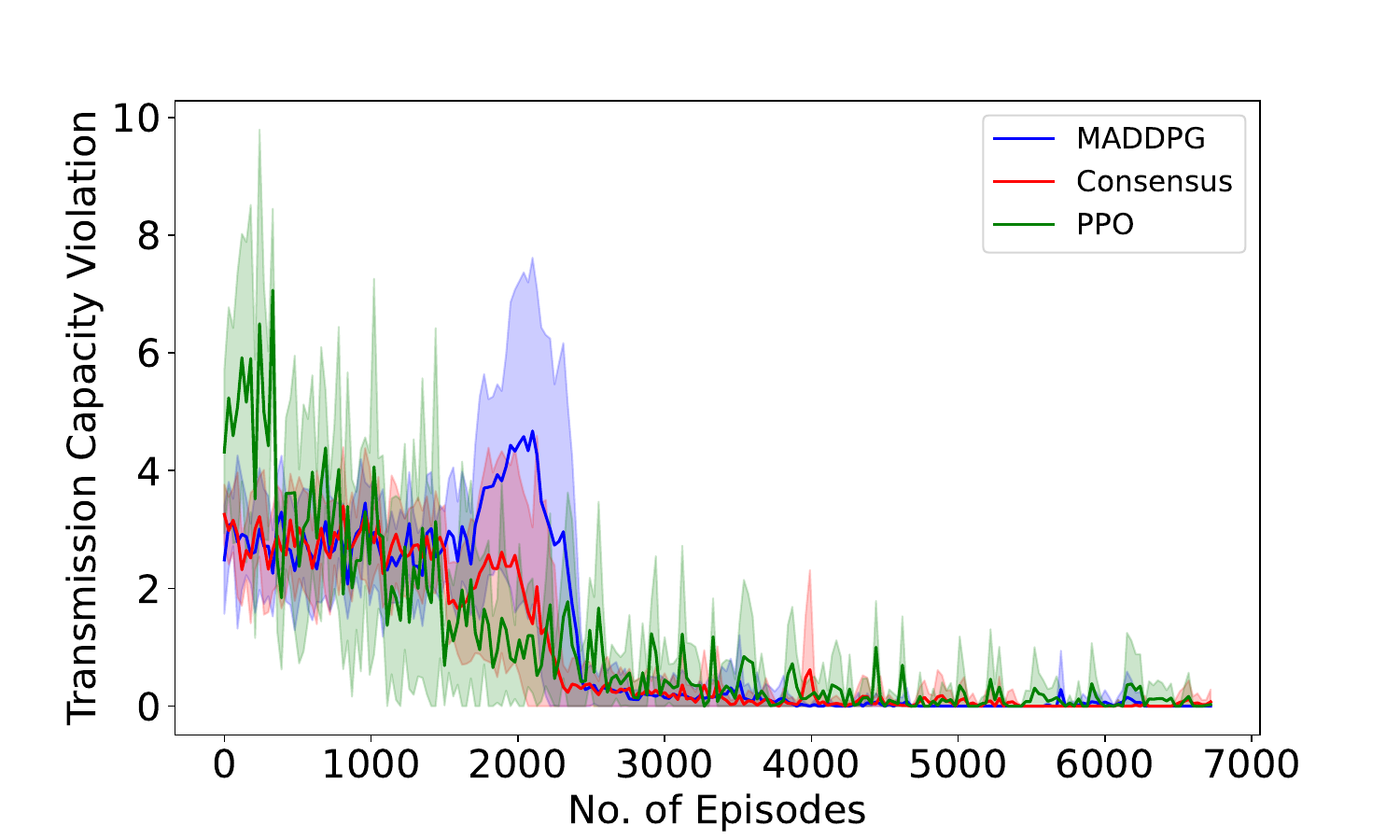}
    \caption{Transmission line capacity violation. The transmission line capacity violation (kVA) in the $n$-th episode is calculated as $\sum_{t=24(n-1)}^{24n-1} \max \{0,S^{684,671}_t-15\}$.}
    \label{fig:capacity}
\end{figure}
Figure~\ref{fig:capacity} illustrates the effectiveness of the three MARL frameworks in reducing distribution line capacity violations. However, the decrease is not always consistent, especially at the beginning of each training cycle for MADDPG. This initial fluctuation can be attributed to the core objective of all MARL frameworks: optimizing the total long-term reward through a balance between exploration (trying new actions) and exploitation (focusing on actions with proven success). During training, certain components of the total reward might temporarily prioritize actions that lead to short-term capacity violations, causing these fluctuations in the observed reduction.



\section{Conclusion and Future Research}
\label{sec:Conclusion}
This study has developed a market and algorithmic framework to enable consumers' and prosumers' participation in local P2P energy trading. Utilizing reinforcement learning algorithms, we have automated bidding for agents while ensuring decentralized decision-making. The SDR-based market clearing addresses challenges with zero-marginal-cost resources and simplifies bidding. Additionally, our MARL framework includes voltage and line capacity constraints of a physical network, setting a foundation for real-world applications. 

However, this research represents just the initial phase in the practical application of a P2P energy trading market, with several challenges ahead. Theoretically, the scalability of the MARL framework is crucial; if the scalability of the consensus MARL algorithm is limited, a mean-field approach, as in \cite{LearningMFG}, could be considered, where agents operate with the belief that the market outcome is at a mean-field equilibrium. Practically, while the current model uses discrete, synchronous trading rounds, future work should investigate continuous and asynchronous trading among agents.


Another critical aspect is cybersecurity. With increased automation, the system becomes vulnerable to cyber attacks, such as malicious users compromising smart inverters to inject false information into the market. In this context, the work in \cite{Gupta}  provides valuable insights. 

Last but not least, it is essential to investigate how short-term P2P market dynamics influence long-term consumer investment decisions, particularly regarding the adoption of solar panels and energy storage systems.

\section*{Acknowledgment}
We acknowledge the support of the U.S. Department of Energy, Office of Electricity under Award Number DE-OE0000921, and the National Science Foundation under Grant No. ECCS-2129631.


\bibliographystyle{elsarticle-num} 
\bibliography{P2PNoMarks}






\end{document}